\documentclass[aps,twocolumn,superscriptaddress,showpacs,pra,amssymb, 
amsmath,latexsym]{revtex4}
\newtheorem{thm}    {Theorem}
\newcommand{\defeq}{\stackrel{\rm def}{=}}

\def\Tr{\mathop{\rm Tr}\nolimits}

\def\integer{\mathbb{Z}}
\def\real{\mathbb{R}}

\def\complex{\mathbb{C}}

\def\SU{\mathop{\rm SU}\nolimits}

\def\U{\mathop{\rm U}\nolimits}
\def\C{\mathop{\rm C}\nolimits}
\def\I{\mathop{\rm I}\nolimits}

\def\Label#1{\label{#1}\ [\ #1\ ]\ }
\def\Label{\label}

\def\GP{\mathop{\rm GP}}
\def\SL{\mathop{\rm SL}}

\def\cl#1{{\cal#1}}
\def\ket#1{|#1\rangle}
\def\bra#1{\langle#1|}

\begin{document}

\title{Discrete realization of group symmetric LOCC-detection 
of maximally entangled state}
\author{Masahito Hayashi}
\email{hayashi@math.is.tohoku.ac.jp}
\address{Graduate School of Information Sciences, Tohoku University, Aoba-ku, Sendai, 980-8579, Japan}
\pacs{03.65.Wj,03.65.Ud,02.20.-a}
\begin{abstract}
Group symmetric LOCC measurement for detecting maximally entangled state
is considered.
Usually, this type measurement has continuous-valued outcomes.
However, any realizable measurement has finite-valued outcomes.
This paper proposes discrete realizations of such a group symmetric LOCC measurement.
\end{abstract}
\maketitle
\section{Introduction}
Testing of maximally entangled state is a useful method for guaranteeing the quality of generated  
maximally entangled states.
However, if we require a group symmetric condition for this method,
the optimal test often requires infinite-valued measurement.
Since any realizable measurement has a finite number of outcomes,
it is needed to discretize 
the optimal measurement.

Now, we focus on the bipartite system ${\cal H}_d \otimes {\cal H}_d$,
in which, the party $A$ and $B$ have the computational bases $\{|i\rangle_A \}_{i=0}^{d-1}$ and $\{|i\rangle_B \}_{i=0}^{d-1}$, respectively.
When our target is testing whether the generated state is sufficiently close to 
the maximal entangled state
\[
	\ket{\phi^0_{AB}}=
	\frac{1}{\sqrt d}\sum_{i=0}^{d-1}
	\ket{i}_{A}\otimes\ket{i}_{B}
\]
under a group symmetric condition, 
the optimal test can be given by
\begin{align}
T_{inv}^{1,A \to B}:=&
\int d
\ket{\varphi \otimes \overline{\varphi}}
\bra{\varphi\otimes \overline{\varphi}}\nu (\,d \varphi)
\nonumber \\
=& \ket{\phi^0_{A,B}}\bra{\phi^0_{A,B}}+
\frac{1}{d+1}(I-\ket{\phi^0_{A,B}}\bra{\phi^0_{A,B}}),\Label{22-17}
\end{align}
where 
$\nu$ is the group invariant probability measure on the set of pure states, and
$\varphi$ and $\overline{\varphi}$ are given as
$\varphi=\sum_{i=0}^{d-1} \varphi_i |i\rangle_A$ 
and $\overline{\varphi}
=\sum_{i=0}^{d-1} \overline{\varphi_i} |i\rangle_B$.
This measurement can be realized by the following procedure.
In the first step, the system $A$ performs the local 
group covariant measurement 
$\int d
\ket{\varphi}
\bra{\varphi}\nu (\,d \varphi)$, and sends the system $B$ the outcome $\varphi$.
In the second step, the system $B$ performs the two-valued measurement
$\{\ket{\overline{\varphi}}\bra{\overline{\varphi}},
I-\ket{\overline{\varphi}}\bra{\overline{\varphi}}\}$.
When Bob obtains the event corresponding to $\{\ket{\overline{\varphi}}\bra{\overline{\varphi}}$,
we support the maximal entangled state $\ket{\phi^0_{AB}}$.

This detection procedure can be generalize as follows.
First, Alice performs a local measurement:
\begin{align*}
M= \{ p_i \ket{u_i}\bra{u_i}\}_i,
\quad \|u_i\|=1 , \quad 0 \le p_i \le 1
\end{align*}
and sends Bob the outcome $i$.
Bob perform the two-valued measurement
$\{
\ket{\overline{u_i}}
\bra{\overline{u_i}},
I-
\ket{\overline{u_i}}
\bra{\overline{u_i}}\} $.
When Bob obtain the event corresponding to 
$\ket{\overline{u_i}}\bra{\overline{u_i}}$,
we support the maximal entangled state $\ket{\phi^0_{AB}}$.
This test can be written as the positive semi-definite matrix $T(M)$:
\begin{align}
T(M)\defeq
\sum_i p_i 
\ket{u_i\otimes \overline{u_i}}
\bra{u_i\otimes \overline{u_i}}.
\end{align}

Indeed, 
when the local dimension $d$ is 2,
D'Ariano et al. \cite{dariano} and Hayashi et al.\cite{tsu}
obtained the discrete own-way LOCC realization of 
the test $T_{inv}^{1,A \to B}$ as test $T(M)$ with an appropriate choice of the local measurement $M$.
However, its general dimensional case was an open problem.
In this paper, 
employing the concepts of 
symmetric informationally complete POVM (SIC-POVM)
and mutually unbiased bases (MUB),
we propose discrete own-way LOCC realizations of $T_{inv}^{1,A\to B}$.
Also, the optimality of the proposed realization scheme is shown.

Next, we consider the case when
Alice's system (Bob's system) is given as 
$\cl{H}_{A_1} \otimes \cl{H}_{A_2}$
($\cl{H}_{B_1} \otimes \cl{H}_{B_2}$)
and the dimensions of all components coincide,
i.e., 
$\dim \cl{H}_{A_1} = \dim \cl{H}_{A_2}= \dim \cl{H}_{B_1} = \dim \cl{H}_{B_2}=d$.
In this case, we focus on
the covariant POVM $M_{cov,u}^2$:
\begin{align*}
& M_{cov,u}^2(\,d g_1\,d g_2)\\
\defeq &
d^2
(g_1\otimes g_2 )\ket{ u }\bra{u}(g_1\otimes g_2 )^*
\nu (\,d g_1)\nu (\,d g_2),
\end{align*}
where the vector $u$ is a maximally entangled state
and
$\nu$ is the group invariant probability measure on $\SU(d)$.
The optimal test is given as the test $T_{inv}^{2,A\to B}\defeq T(M_{cov,u}^2)$,
which has the form \cite{Haya}:
\begin{align}
& T_{inv}^{2,A\to B}\nonumber \\
=& \ket{\phi^0_{A,B}}\bra{\phi^0_{A,B}}
\otimes
\ket{\phi^0_{A,B}}\bra{\phi^0_{A,B}}\nonumber\\
& + 
\frac{1}{d^2-1}(I-\ket{\phi^0_{A,B}}\bra{\phi^0_{A,B}})
\otimes (I-\ket{\phi^0_{A,B}}\bra{\phi^0_{A,B}}).
\end{align}
Indeed, 
The positive semi-definite matrix $T_{inv}^{2,A\to B}$
does not depend on the choice of the maximally entangled state $u$.
In this paper, 
employing the concept of Clifford group,
we provide a 
discrete own-way LOCC realization of $T_{inv}^{2,A\to B}$
when the local system is given as a composite system of 
a prime-dimensional system.
Also, the optimality of the proposed realizaion scheme is shown.

\section{Discrete own-way LOCC realization of $T_{inv}^{1,A\to B}$}
\subsection{Realizing scheme by SIC-POVM}
In order to design the test $T_{inv}^{1,A\to B}$,
we focus on the concept ``symmetric informationally complete POVM 
(SIC-POVM)''.
A rank-one POVM $\{p_i \ket{u_i}\bra{u_i}\}$ on 
$\cl{H}_A= \complex^d$ is called 
a {\it symmetric informationally complete POVM (SIC-POVM)}, if 
it satisfies the following conditions:
\begin{align}
\#\{i\}=& d^2,\nonumber \\
p_i= & \frac{1}{d}\nonumber \\
| \langle u_i | u_j \rangle |^2= & \frac{1}{d+1} \hbox{ for }i \neq j
\Label{2-2-4}
\end{align}
Currently, an SIC-POVM analytically 
is constructed when the dimension $d$ is 
2,3\cite{KK,Z},4\cite{RBSC,Z},5\cite{Z},6\cite{G},7\cite{A},
8\cite{KK}, or 19\cite{A}.
Also, its existence is numerically verified up to $d=45$\cite{RBSC}.
As is shown in Appendix \ref{a15},
any SIC-POVM $M_{sic}=\{p_i \ket{u_i}\bra{u_i}\}_i$
satisfies
\begin{align}
T(M_{sic})=T_{inv}^{1,A\to B}, \Label{2-2-6}
\end{align}
that is, the test $T_{inv}^{1,A\to B}$ can be realized by
an SIC-POVM.
Moreover,
if a POVM $M=\{M_i\}_i$ on $\cl{H}_A$ satisfies 
\begin{align*}
T(M)= T_{inv}^{1,A\to B},
\end{align*}
the inequality
\begin{align*}
\#\{i\}\ge d^2
\end{align*}
holds. This is because 
the rank of the operator $T_{inv}^{1,A\to B}$ (which equal $d^2$)
is less than the number of the elements of POVM $M_i$.
Hence, we obtain
\begin{align}
\min\{\#\{i\}| T(\{M_i\}_i)= T_{inv}^{1,A\to B}\}= d^2
\label{18-1}
\end{align}
if there exists an SIC-POVM on $\complex^d$.
That is, the proposed realizing scheme by SIC-POVM is optimal in the sense of (\ref{18-1}).

\subsection{Realizing scheme by MUB}
However, any SIC-POVM is not 
a randomized combination of projection valued measures
as well as a projection valued measure.
Since a projection valued measure (PVM) are more realizable than other POVM,
it is more desired to design Alice's POVM as
a randomized combination of PVMs.
For this purpose, we focus on mutually unbiased bases.
$d+1$ orthonormal bases $\{\cl{B}_1, \ldots, \cl{B}_{d+1}\}$ 
are called mutually unbiased bases (MUB)
if 
\begin{align*}
| \langle u | v \rangle |^2= \frac{1}{d}, \forall
u \in \cl{B}_i, \forall v \in \cl{B}_j, i \neq j.
\end{align*}
The existence of MUB is shown when $d$ is a prime\cite{Ivan} or 
a prime power\cite{WF}.
Bandyopadhyay {\it et~al.} gave a more explicit form in these cases
\cite{BBRV}.
Any mutually unbiased bases
$\{\cl{B}_1, \ldots, \cl{B}_{d+1}\}$
make the POVM $M_{\cl{B}_1, \ldots, \cl{B}_k}$, {\it i.e.},
\begin{align*}
M_{\cl{B}_1, \ldots, \cl{B}_{d+1}}= \left\{
\frac{1}{d+1}
\ket{u_{i,j}}\bra{u_{i,j}}
\right\}_{i,j},
\end{align*}
where $\cl{B}_j= \{ u_{1,j}, \ldots, u_{d,j}\}$.
This POVM always produces the desired test $T_{inv}^{1,A\to B}$ as
\begin{align}
T(M_{\cl{B}_1, \ldots, \cl{B}_{d+1}})=T_{inv}^{1,A\to B},\Label{2-2-7}
\end{align}
which is shown in Appendix \ref{a16}.
This construction of the test $T_{inv}^{1,A\to B}$ is 
optimal in the following sense.
Let $\{M^j\}$ be the set of projection-valued measures.
A randomized combination of $\{M^j\}$, {\it i.e.},
$M= \sum_j p_j M_j$
satisfies  
\begin{align}
T(M)= T_{inv}^{1,A\to B}.\Label{2-3-1}
\end{align}
Then,
as is proven in Appendix \ref{a17},
\begin{align}
\#\{j\} \ge d+1, \Label{2-2-1}
\end{align}
which implies the optimality of 
the POVM consisting of MUB.
Hence, 
\begin{align}
\min_{M_j: \hbox{\rm \tiny PVM}}\left\{ \#\{j\}\left|
T\left(\sum p_j M_j\right)= T_{inv}^{1,A\to B}
\right. \right\}= d+1\label{18-2}
\end{align}
if $d$ is a prime or a prime power.
That is, the proposed realizing scheme by MUB is optimal in the sense of (\ref{18-2}).

\section{Discrete own-way LOCC realization of $T_{inv}^{2,A\to B}$}
Next, we proceed to the case when 
both local systems consist of two subsystems. 
Given a finite group $G$ and its projective representation 
$f$ on $\cl{H}_{A_1} = \complex^d$,
by regarding $\cl{H}_{A_2}$ as the dual space of $\cl{H}_{A_1}$,
the matrix $f(g)$ can be regarded as
an element $|f(g) \rangle $ of $\cl{H}_{A_1} \otimes \cl{H}_{A_2}$.

\begin{thm}\label{thm}
We assume the two conditions:
(1) The representation $f$ is irreducible.
(2) The action $f \otimes \overline{f}$ of $G$ to 
$\cl{H}_{A_1} \otimes \cl{H}_{A_2}$ has only two irreducible components,
{\it i.e.,}
the irreducible subspaces of $\cl{H}_{A_1} \otimes \cl{H}_{A_2}$
for the action 
\begin{align*}
v_1\otimes v_2 \to 
f(g) v_1\otimes \overline{f(g)} v_2 
\end{align*}
are only
the one-dimensional space $<\phi_{A_1,A_2}^0>$ and its orthogonal space 
$<\phi_{A_1,A_2}^0>^{\perp}$.
Then,
the resolution $
M_{f}= 
\left\{ \frac{d^2}{|G|}\left\ket{\frac{1}{\sqrt{d}}f(g)\right}
\left\bra{\frac{1}{\sqrt{d}}f(g)\right}\right\}
_{g \in G}$ satisfies the condition for a POVM,
and 
\begin{align}
T(M_{f})
=  T_{inv}^{2,A\to B}.\Label{2-2-2}
\end{align}
\end{thm}
Its proof is given in Appendix \ref{a20}.
This theorem yields 
a discrete own-way LOCC realization of $T_{inv}^{2,A\to B}$
from the representation $f$ satisfying the above two conditions.

For example, Clifford group satisfies this assumption.
For readers' convenience, we give its definition and 
prove that Clifford group satisfies this assumption.
Clifford group $\C(d)$ for $d$-dimensional system
is given by
\begin{align*}
\C(d) &:= \{U\in \U(d)|
U \GP(d)U^\dagger= \GP(d)
\}\\
\GP(d) &:=
\{e^{\sqrt{-1} \xi} W(i,j)| \xi \in \real, i,j \in  \integer \} \\
\I(d)&:= 
\{e^{\sqrt{-1} \xi} | \xi \in \real\},
\end{align*}
where
\begin{align*}
Z&:= \sum_{j=0}^{d-1} \omega^{j} |j\rangle \langle j| , \quad
X:= \sum_{j=0}^{d-1} |j+1\rangle \langle j| \\
W(i,j)&:=X^i Z^j 
\end{align*}
and $\omega$ is the $d$-th root of $1$.
As is shown in Appendix \ref{a18}, 
the natural representation of the group $\C(d)$ satisfies the conditions (1) and (2).
Then, the natural projective representation of the group $\C(d)/\I(d)$ also satisfies the conditions (1) and (2).
As is shown in Lemma  5 in Appleby \cite{A}, when $d$ is prime, 
the cardinality $|\C(d)/\I(d)|$ is $d^3(d^2-1)$.
In the general case, 
\begin{align*}
|\C(d)/\I(d)|
=d^2 
\left(
\sum_{n=0}^{d-1}\nu(n,d)\nu(n+1,d)
\right),
\end{align*}
where $\nu(n,d)$ is the number of distinct ordered pairs $(x, y) \in \integer_d^2$
such that $xy = n$ (mod $d$).

\section{Discussion}
This paper has treated discretization of onw-way LOCC protocols.
Using the concepts of 
symmetric informationally complete POVM (SIC-POVM), 
mutually unbiased bases (MUB),
and Clifford group,
we have proposed discrete own-way LOCC realizations of 
$T_{inv}^{1,A\to B}$ and $T_{inv}^{2,A\to B}$.
This result indicates the importance of these concept in discrete mathematics.
Since the existence of SIC-POVM and MUB is proven in limited cases,
we cannot construct a discrete own-way LOCC realization of 
$T_{inv}^{1,A\to B}$ in the general case.
Thus, further investigation for these concepts are required.

While the optimal test is given as $T_{inv}^{3,A\to B}$
when the local system consists of three subsystems by Hayashi \cite{Haya},
its discretization has not been obtained.
Since the optimal test $T_{inv}^{3,A\to B}$ is closely related to GHZ state\cite{Haya},
its discretization may be related to GHZ state.
Its construction remains as a future research.

Further, 
the optimal protocol is often given as a protocol with infinite elements
in quantum information.
In such a case, it is required to discretize this protocol.
This kind of discretization is 
an interesting interdisciplinary topic between 
quantum information and discrete mathematics.

\section*{Acknowledgment}
This research
was partially supported by a Grant-in-Aid for Scientific Research on Priority Area `Deepening and Expansion of Statistical Mechanical Informatics (DEX-SMI)', No. 18079014 and
a MEXT Grant-in-Aid for Young Scientists (A) No. 20686026.

\appendix
\section{Proof of (\ref{2-2-6})}\Label{a15}
First, we show that
$u_1 \otimes \overline{u_1}, \ldots, u_{d^2}\otimes\overline{u_{d^2}}$ 
are linearly independent.
We choose complex numbers $a_1, \ldots, a_{d^2}$
such that
\begin{align*}
\sum_i a_i u_i\otimes \overline{u_i}= 0.
\end{align*}
Taking trace, we have
\begin{align*}
a_ 1 + \sum_{i\neq 1} a_i = 0 .
\end{align*}
On the other hand,
\begin{align*}
0= \langle u_1\otimes \overline{u_1}  
\ket{\sum_i a_i u_i \otimes \overline{u_i}}
= a_1 + \frac{1}{d+1}\sum_{i \neq 1}a_i.
\end{align*}
Hence, we obtain $a_1=0$.
Similarly, we can show $a_i=0$,
which implies the linear independence.

Since the dimension of $\cl{H}_A\otimes \cl{H}_B$ is $d^2$,
any element of $\cl{H}_A\otimes \cl{H}_B$ can be expressed as
\begin{align*}
\sum_j 
a_i u_i\otimes \overline{u_i}.
\end{align*}
We can calculate
\begin{align*}
& 
\left\bra{\sum_i  a_i u_i\otimes \overline{u_i}\right}
T(M_{sic})
\left\ket{\sum_j  a_j u_j\otimes \overline{u_j}\right} \\
= &
\left\bra{\sum_i  a_i u_i\otimes \overline{u_i}\right}
\left(\sum_k
\frac{1}{d}
\left\ket{u_k \otimes \overline{u_k}\right}
\left\bra{u_k \otimes \overline{u_k}\right}
\right)
\left\ket{\sum_j  a_j u_j\otimes \overline{u_j}\right} \\
= &
\frac{d+2}{(d+1)^2}
\left|\sum_k a_k\right|^2 
+ \frac{d}{(d+1)^2}
\sum_k |a_k|^2 .
\end{align*}
On the other hand,
its norm is calculated as
\begin{align*}
\left\|
\sum_j  a_j u_j\otimes \overline{u_j}
\right\|= 
\frac{1}{d+1}
\left|\sum_k a_k\right|^2 
+ \frac{d}{d+1}\sum_k |a_k|^2 .
\end{align*}
Since 
\begin{align*}
\left|\langle \phi_{A,B}^0
\ket{\sum_j  a_j u_j\otimes \overline{u_j}}\right|^2
= \frac{1}{d} \left|\sum_j  a_j\right|^2,
\end{align*}
we obtain
\begin{align*}
& 
\left\bra{\sum_i  a_i u_i\otimes \overline{u_i}\right}
T_{inv}^{1,A\to B}
\left\ket{\sum_j  a_j u_j\otimes \overline{u_j}\right} \\
= &
\left\bra{\sum_i  a_i u_i\otimes \overline{u_i}\right}
\left(\frac{d}{d+1}  \ket{\phi_{A,B}^0}\bra{\phi_{A,B}^0}
+ \frac{1}{d+1}I\right)\\
& ~\left\ket{\sum_j  a_j u_j\otimes \overline{u_j}\right} \\
=&
\frac{d}{d+1}\frac{1}{d}
\left|\sum_j  a_j\right|^2\\
& +
\frac{1}{d+1}
\left(
\frac{1}{d+1}
\left|\sum_k a_k\right|^2 
+ \frac{d}{d+1}\sum_k |a_k|^2 
\right)\\
= &
\left\bra{\sum_i  a_i u_i\otimes \overline{u_i}\right}
T(M_{sic})
\left\ket{\sum_j  a_j u_j\otimes \overline{u_j}\right} .
\end{align*}
Therefore, we obtain (\ref{2-2-6}).

\section{Proof of (\ref{2-2-7})}\Label{a16}
We focus on the subspace $<\phi_{A,B}^0>^\perp$ orthogonal to
$\phi_{A,B}^0$.
The subspace $\cl{B}'_j=
<u_{1,j}\otimes \overline{u_{1,j}}-\frac{1}{d} \phi_{A,B}^0,
\ldots,
u_{d-1,j}\otimes \overline{u_{d-1,j}}-\frac{1}{d} \phi_{A,B}^0>$
belongs to the subspace $<\phi_{A,B}^0>^\perp$,
and its dimension is $d-1$.
Since
\begin{align}
\langle 
u_{i,j}\otimes \overline{u_{i,j}}-\frac{1}{d} \phi_{A,B}^0|
u_{i',j'}\otimes \overline{u_{i',j'}}-\frac{1}{d} \phi_{A,B}^0
\rangle=0 , \quad j\neq j',
\end{align}
The spaces $\cl{B}'_1, \ldots, \cl{B}'_{d+1}$ are
orthogonal to each other.
Since the dimension of the subspace $<\phi_{A,B}^0>^\perp$ is $d^2-1$,
the subspace $<\phi_{A,B}^0>^\perp$ 
is spanned by the spaces $\cl{B}'_1, \ldots, \cl{B}'_{d+1}$.
Therefore,
any element of the space $\cl{H}_A\otimes \cl{H}_B$ 
can be expressed as
$\sum_{j=1}^{d+1}\sum_{i=1}^d 
a_{i,j} u_{i,j}\otimes \overline{u_{i,j}}$.
In the following, we abbreviate the sum 
$\sum_{j=1}^{d+1}\sum_{i=1}^d $ as $\sum_{j,i}$.

We calculate
\begin{align*}
& \left\bra{
\sum_{j,i}
a_{i,j} u_{i,j}\otimes \overline{u_{i,j}}
\right}
T(M_{\cl{B}_1, \ldots, \cl{B}_{d+1}})
\left\ket{\sum_{j',i'}
a_{i',j'} u_{i',j'}\otimes \overline{u_{i',j'}}\right} 
\\
= &
\left\bra{
\sum_{j,i}
a_{i,j} u_{i,j}\otimes \overline{u_{i,j}}
\right}
\left(\sum_{l,k}
\frac{1}{d+1}
\ket{u_{k,l} \otimes \overline{u_{k,l}}}
\bra{u_{k,l} \otimes \overline{u_{k,l}}}
\right)\\
& \hspace{10ex}
\left\ket{\sum_{j',i'}
a_{i',j'} u_{i',j'}\otimes \overline{u_{i',j'}}\right} 
\\
= &
\frac{1}{d+1}
\sum_{l,k}
\left|
\sum_{j,i}
\bra{u_{k,l} \otimes \overline{u_{k,l}}}
\ket{
a_{i,j} u_{i,j}\otimes \overline{u_{i,j}}} 
\right|^2\\
= &
\frac{1}{d}
\left|\sum_{j,i}a_{i,j}\right|^2
-
\frac{1}{d(d+1)}
\sum_j | \sum_i a_{i,j}|^2
+ \frac{1}{d+1}
\sum_{j,i} | a_{i,j}|^2
\end{align*}
On the other hand,
its norm is calculated as
\begin{align*}
&\left\|
\sum_{j,i}
a_{i,j} u_{i,j}\otimes \overline{u_{i,j}}
\right\|\\
=& 
\frac{1}{d}
\left|\sum_{j,i}a_{i,j}\right|^2
-
\frac{1}{d}
\sum_j | \sum_i a_{i,j}|^2
+ 
\sum_{j,i} | a_{i,j}|^2
\end{align*}
Since 
\begin{align*}
\left|\left\langle \phi_{A,B}^0\left|
\sum_{j,i}
a_{i,j} u_{i,j}\otimes \overline{u_{i,j}}\right.\right\rangle
\right|^2
= \frac{1}{d} \left|a_{i,j} u_{i,j}\right|^2,
\end{align*}
we obtain
\begin{align*}
& 
\left\bra{
\sum_{j,i}
a_{i,j} u_{i,j}\otimes \overline{u_{i,j}}
\right}
T_{inv}^{1,A\to B}
\left\ket{\sum_{j',i'}
a_{i',j'} u_{i',j'}\otimes \overline{u_{i',j'}}\right} 
\\
= &
\left\bra{
\sum_{j,i}
a_{i,j} u_{i,j}\otimes \overline{u_{i,j}}
\right}
\frac{d}{d+1}  \ket{\phi_{A,B}^0}\bra{\phi_{A,B}^0}
+ \frac{1}{d+1}I\\
& \hspace{10ex}\left\ket{\sum_{j',i'}
a_{i',j'} u_{i',j'}\otimes \overline{u_{i',j'}}\right} \\
=&
\frac{1}{d}
\left|\sum_{j,i}a_{i,j}\right|^2
-
\frac{1}{d(d+1)}
\sum_j | \sum_i a_{i,j}|^2
+ \frac{1}{d+1}
\sum_{j,i} | a_{i,j}|^2\\
= & 
\left\bra{
\sum_{j,i}
a_{i,j} u_{i,j}\otimes \overline{u_{i,j}}
\right}
T(M_{\cl{B}_1, \ldots, \cl{B}_{d+1}})
\left\ket{\sum_{j',i'}
a_{i',j'} u_{i',j'}\otimes \overline{u_{i',j'}}\right} .
\end{align*}
Therefore, we obtain (\ref{2-2-7}).

\section{Proof of (\ref{2-2-1})}\Label{a17}
Let $M^j= \{ \ket{u_{i,j}}\bra{u_{i,j}}\}$.
We focus on the projection $P$ to 
the subspace $<\phi_{A,B}^0>^\perp$ orthogonal to
$\phi_{A,B}^0$ and 
the subspace 
$\cl{B}''_j\defeq <u_{1,j}\otimes \overline{u_{1,j}},
\ldots,
u_{d,j}\otimes \overline{u_{d,j}}>$.
The image $P \cl{B}''_j$ 
is $<u_{1,j}\otimes \overline{u_{1,j}}-\frac{1}{d} \phi_{A,B}^0,
\ldots,
u_{d-1,j}\otimes \overline{u_{d-1,j}}-\frac{1}{d} \phi_{A,B}^0>$.
The condition (\ref{2-3-1}) implies
that the sum of the rank of the space $P \cl{B}''_j$ 
is greater than $d^2-1$, {\it i.e.},
the dimension of the space $<\phi_{A,B}^0>^\perp$.
Thus,
$\#\{j\} (d-1)\ge d^2-1$,
which implies the inequality (\ref{2-2-1}).

\section{Proof of Theorem \ref{thm}}\Label{a20}
First, we prove that $M_{f}$ satisfies the condition for POVM.
The irreducibility of the action $f$ guarantees that
\begin{align*}
& \frac{d}{|G|}
\sum_{g \in G}
\bra{k}f(g)\ket{l}\bra{l'}f(g)\ket{k'}\\
=&
\bra{k}
\left(\frac{d}{|G|}
\sum_{g \in G}
f(g)\ket{l}\bra{l'}f(g)\right)
\ket{k'}\\
=&\bra{k}
\langle l|l'\rangle I
\ket{k'}
=\delta_{k,k'}\delta_{l,l'}.
\end{align*}

we obtain
\begin{align*}
&\frac{d}{|G|}
\sum_{g \in G}
\left|
\langle f(g)|
\left(
\sum_{k,l} a_{k,l} 
|k \rangle \otimes |l \rangle
\right)
\right|^2 
\\
=&\frac{d}{|G|}\sum_{g \in G}
\sum_{k,l}\sum_{k',l'}
a_{k,l} \overline{a_{k',l'}}
\overline{\bra{k}f(g)\ket{l}}
{\bra{l'}f(g)\ket{k'}}
\\
= &\sum_{k,l}a_{k,l} \overline{a_{k,l}},
\end{align*}
which implies
\begin{align*}
\frac{d^2}{|G|}
\sum_{g \in G}
\left\ket{\frac{1}{\sqrt{d}}f(g)\right}
\left\bra{\frac{1}{\sqrt{d}}f(g)\right}
= I_{{A_1},{A_2}}.
\end{align*}
Hence, $
M_{f}= 
\left\{ \frac{d^2}{|G|}\left\ket{\frac{1}{\sqrt{d}}f(g)\right}
\left\bra{\frac{1}{\sqrt{d}}f(g)\right}\right\}
_{g \in G}$
is a POVM.

Next, we show (\ref{2-2-2}).
We focus on the action of the group $G \times G$
to the total space
$\cl{H}_{A_1}\otimes \cl{H}_{A_2}\otimes \cl{H}_{B_1}\otimes 
\cl{H}_{B_2}$ as
\begin{align*}
&u_1\otimes u_2 \otimes v_1\otimes v_2 \\
&\quad \mapsto 
f(g_1) u_1\otimes \overline{f(g_2)} u_2 \otimes 
\overline{f(g_1)}v_1\otimes f(g_2) v_2 \\
\end{align*}
for 
$u_i \in \cl{H}_{A_i}$,
$v_i \in \cl{H}_{B_i}$, and 
any pair $(g_1,g_2) \in G \times G$.
Due to the condition (2), 
the irreducible decomposition of the space
$\cl{H}_{A_1}\otimes \cl{H}_{A_2}\otimes \cl{H}_{B_1}\otimes 
\cl{H}_{B_2}$
is given as
$<\phi_{A_1,B_1}^0>\otimes <\phi_{A_2,B_2}^0>
\oplus <\phi_{A_1,B_1}^0>\otimes <\phi_{A_2,B_2}^0>^{\perp}
\oplus <\phi_{A_1,B_1}^0>^{\perp}\otimes <\phi_{A_2,B_2}^0>
\oplus <\phi_{A_1,B_1}^0>^{\perp}\otimes <\phi_{A_2,B_2}^0>^{\perp}$.

As is checked below, the test $T(M_f)$ is invariant for this action:
\begin{widetext}
\begin{align*}
& f(g_1) \otimes \overline{f(g_2)} 
\otimes \overline{f(g_1)}
\otimes f(g_2)
T(M_f)
\left(
f(g_1) \otimes \overline{f(g_2)} 
\otimes \overline{f(g_1)}
\otimes f(g_2)
\right)^{\dagger}\\
=& 
\frac{d^2}{|G|}
\sum_{g \in G} 
\left\ket{
\frac{1}{d}f(g_1) f(g)f(g_2)^\dagger \otimes 
\overline{f(g_1)}
\overline{f(g)}
\overline{f(g_2)^\dagger}
\right}
\left\bra{
\frac{1}{d}f(g_1) f(g)f(g_2)^\dagger \otimes 
\overline{f(g_1)}
\overline{f(g)}
\overline{f(g_2)^\dagger}
\right}\\
=& 
\frac{d^2}{|G|}
\sum_{g' \in G} 
\left\ket{
\frac{1}{d}f(g')\otimes 
\overline{f(g')}
\right}
\left\bra{
\frac{1}{d}f(g')\otimes 
\overline{f(g')}
\right}
= T(M_f),
\end{align*}
\end{widetext}
where we denote $g_1 g g_2^{-1}$ by $g'$.
Hence, the test $T(M_f)$ has the form
\begin{align*}
&T(M_f)\\
=& 
a\ket{\phi_{A_1,B_1}^0}\bra{\phi_{A_1,B_1}^0}
\otimes \ket{\phi_{A_2,B_2}^0}\bra{\phi_{A_2,B_2}^0}\\
& +
b
(I- \ket{\phi_{A_1,B_1}^0}\bra{\phi_{A_1,B_1}^0})
\otimes \ket{\phi_{A_2,B_2}^0}\bra{\phi_{A_2,B_2}^0} \\
& + 
c
\ket{\phi_{A_1,B_1}^0}\bra{\phi_{A_1,B_1}^0}
\otimes (I- \ket{\phi_{A_2,B_2}^0}\bra{\phi_{A_2,B_2}^0})\\
& +
 d
(I- \ket{\phi_{A_1,B_1}^0}\bra{\phi_{A_1,B_1}^0})
\otimes (I- \ket{\phi_{A_2,B_2}^0}\bra{\phi_{A_2,B_2}^0}).
\end{align*}
Since $f(g)$ is the unitary matrix,
$\frac{1}{\sqrt{d}}f(g)$ is a maximally entangled state
on $\cl{H}_{A_1}\otimes \cl{H}_{A_2}$.
Since 
$\left\ket{\frac{1}{\sqrt{d}}f(g)\right}$
is maximally entangled,
Lemma 5 in Hayashi \cite{Haya} yields that
\begin{align}
T(M_f)= 
 \ket{\phi^0_{A_1,B_1}\otimes \phi^0_{A_2,B_2}}
\bra{\phi^0_{A_1,B_1}\otimes \phi^0_{A_2,B_2}}
+
P T(M_f) P,
\end{align}
where
\begin{align*}
P\defeq (I- \ket{\phi^0_{A_2,B_2}}\bra{\phi^0_{A_2,B_2}})
\otimes (I- \ket{\phi^0_{A_1,B_1}}\bra{\phi^0_{A_1,B_1}}).
\end{align*}
This relation implies that $b=c=0$.
Thus, the relation $\Tr T(M_f)= d^2$ yields
\begin{align*}
&T(M_f)\\
=& 
\ket{\phi_{A_1,B_1}^0}\bra{\phi_{A_1,B_1}^0}
\otimes \ket{\phi_{A_2,B_2}^0}\bra{\phi_{A_2,B_2}^0}\\
& \quad +
\frac{1}{d^2-1}
(I- \ket{\phi_{A_1,B_1}^0}\bra{\phi_{A_1,B_1}^0})
\otimes (I- \ket{\phi_{A_2,B_2}^0}\bra{\phi_{A_2,B_2}^0}),
\end{align*}
which implies (\ref{2-2-2}).

\section{Proof of irreducibility}\Label{a18}
It is known that 
the natural representation of
the subgroup $\GP(d)\subset \C(d)$ satisfies the condition (1).
Hence, it is sufficient to show the condition (2).
the irreducible spaces of the subgroup $\GP(d)\subset \C(d)$ are
$d^2$ one-dimensional subspaces generated by $| W(i,j)\rangle $ for $i,j$.
The representation of $\GP(d)$ on each irreducible subspaces is different.
Thus, the irreducible subspace of the larger group $\C(d)$ should be represented as
the direct sum of these subspaces.
As is shown in Lemma 1 in Appleby \cite{A},
for any $(i,j)$ and any $F\in \SL(2,\integer_{\overline{d}})$,
there exists an element $U \in \C(d)$ such that
$f(U)\otimes \overline{f(U)}
| W(i,j)\rangle = e^{\sqrt{-1} \delta_{i,j,,F}}
| W(F(i,j))\rangle$,
where
\begin{align}
\overline{d}:=
\left\{
\begin{array}{ll}
d & \hbox{if} d \hbox{is odd}\\
2d & \hbox{if} d \hbox{is even} .
\end{array}
\right.
\end{align}
For any pair $(i,j)=\neq (0,0)$, there exists an element $F \in \SL(2,\integer_{\overline{d}})$
such that $(i,j)=F(1,0)$.
Since any irreducible subspace should be spanned by the subset of 
$\{| W(i,j)\rangle\}_{i,j}$,
the space spanned by $\{| W(i,j)\rangle\}_{(i,j)\neq (0,0)}$
is irreducible.
Thus, the condition (2) holds.

\end{document}